**Imaging through nonlinear metalens using second harmonic generation**

*Christian Schlickriede, Naomi Waterman, Bernhard Reineke, Philip Georgi, Guixin Li, Shuang Zhang, Thomas Zentgraf\**

C. Schlickriede, Bernhard Reineke, Philip Georgi, Prof. T. Zentgraf
Department of Physics
University of Paderborn
Warburger Straße 100, D-33098, Paderborn, Germany
Email: thomas.zentgraf@uni-paderborn.de

G. Li
Department of Materials Science and Engineering,
Southern University of Science and Engineering,
Shenzhen, 518055, China
Email: ligx@sustc.edu.cn

N. Waterman, Prof. S. Zhang
School of Physics & Astronomy
University of Birmingham
Birmingham, B15 2TT
Email: s.zhang@bham.ac.uk

**The abrupt phase change of light at metasurfaces provides high flexibility in wave manipulation without the need of accumulation of propagating phase through dispersive materials. In the linear optical regime, one important application field of metasurfaces is imaging by planar metalenses, which enables device miniaturization and aberration correction compared to conventional optical microlens systems. With the incorporation of nonlinear responses into passive metasurfaces, optical functionalities of metalenses are anticipated to be further enriched, leading to completely new applications areas. Here, we demonstrate imaging with nonlinear metalenses that combine the function of an ultrathin planar lens with simultaneous frequency conversion. With such nonlinear metalenses, we experimentally demonstrate imaging of objects with near infrared light while the image appears in the second harmonic signal of visible frequency range. Furthermore, the functionality of these nonlinear metalenses can be modified by switching the handedness of the circularly polarized fundamental wave, leading to either real or virtual nonlinear image formation. Nonlinear metalenses do not only enable infrared light imaging through a visible detector but also have the ability to modulate nonlinear optical responses through an ultrathin metasurface device while the fundamental wave remains unaffected, which offers the capability of nonlinear information processing with novel optoelectronic devices.**

Conventional optical focusing, which can be classified by refractive and diffractive effects, is limited either by the surface topography or the refractive index contrast dictated by the materials or the fabrication processes. Metamaterials have shown the potential to relax some constraints in the control of light propagation, enabling new or improved functionalities such as invisibility cloaking[1] and sub-diffraction-limited super-imaging.[2] A sophisticated manipulation of both phase and amplitude of light[3] can be realized by designing the geometries and orientation angles of meta-atoms on planar metasurfaces.[4-9] In this way, both plasmonic and dielectric metalenses with diverse optical functionalities have been demonstrated.[10-14] One example are dual-polarity metalenses[7, 15] that act as either convex or concave lenses depending on the circular polarization state of the light. The underlying mechanism of such dual-polarity lenses results from a spin-dependent Pancharatnam-Berry (P-B) phase.[16-19] Meanwhile, a plethora of optical effects using P-B phase metasurfaces have been successfully demonstrated, including three-dimensional optical holography,[8, 20, 21] generation of vortex beams,[5, 22] laser beam shaping,[23, 24] multifocal lenses,[25] integrated dynamical phase lenses[26, 27] or helicity-dependent multifunctional lenses.[28, 29] The concept of geometric P-B phases has been recently applied for tailoring the nonlinear optical responses of plasmonic metasurfaces.[30] Nonlinear spin-orbit interactions[31] and nonlinear optical holography[32] have been demonstrated through harmonic generation processes. Plasmon-enhanced nonlinear processes were also proposed for superresolution applications.[33] While the conversion efficiency of such nonlinear devices is generally too low for practical applications, Lee et al. demonstrate that the second harmonic generation (SHG) can be strongly enhanced by coupling the plasmonic resonances of the meta-atoms with intersubband transitions in semiconductor quantum wells, paving the way for high-efficiency nonlinear metasurfaces.[34] There are other strategies to increase the efficiency, for instance, using different material platforms. Strong third harmonic generation (THG) has also been introduced from Fano-resonant Si-based metasurfaces[35] leading to an enhancement factor of $> 10^5$ compared to bulk material. Gallium arsenide based dielectric metasurfaces for SHG are reported, which show an enhancement factor of $10^4$ compared to the unstructured film.[36] The advantage by using nonlinear optics with resonant dielectric nanostructures is that it grants a 3D volume that is not restricted to interfaces and thus may lead to higher conversion efficiencies due to Fabry-Perot effects. However, so far there is no dielectric high-efficiency metasurface for the nonlinear regime, which is able to spatially control the phase at the same time. Meanwhile, nonlinear plasmonic metalenses based on four-wave mixing and third harmonic generation have been introduced.[37, 38] However, those metalenses rely on the dispersion control of the individual meta-atoms, which introduces more complexities and phase uncertainties in the design and device fabrication. Furthermore, the question remains whether such nonlinear metalenses can be used for nonlinear optical imaging of objects and not only for relatively simple focusing of Gaussian beams.

Here, we design and experimentally demonstrate a nonlinear metalens for nonlinear light focusing and imaging. The nonlinear metalens acts like a regular lens but with the difference that it only controls the beam propagation of the nonlinear signal, which is simultaneously generated when the light from the object interacts with the metalens. The propagation of the fundamental light remains unaltered. For our demonstration, we are using the concept of the nonlinear P-B phase for SHG from meta-atoms with three-fold (C3) rotational symmetry. Based on the symmetry selection rules,[39, 40] SHG from a C3 meta-atom has opposite circular polarization compared to that of the fundamental wave. By controlling the orientation angle of the meta-atom, the local phase of SHG can be continuously controlled over the entire $2\pi$ range. By assembling the C3 meta-atoms into a configuration that resembles the phase profile of a lens for the SHG wavelength, we demonstrate helicity dependent nonlinear beam focusing and nonlinear imaging by using second harmonic generations at visible wavelengths.

The concept of nonlinear metalens is schematically shown in Figure 1a. Light from an object at the fundamental wavelength is collected by the metalens, partially converted to the second harmonic wavelength and imaged to an image plane, which is determined by the focal length and the distance of the object to the lens. The functionality of an optical lens can be realized by a suitable spatially variant phase that is simultaneously added to the nonlinear signal generated at each meta-atom on the surface. For a meta-atom with C3 rotational symmetry, a nonlinear phase of $3\sigma\theta$ is introduced to the SHG wave, where σ=±1 represents the left and right circular polarizations (LCP and RCP) and $\theta$ is the orientation angle of the meta-atom.[31] The required spatial phase distribution for a lens is given by

$$\varphi(r) = k_0 \left(\sqrt{f^2 + r^2} - |f|\right), \qquad (1)$$

where r is the radial distance from the lens axis of the metalens, $f$ is the focal length of the lens in air and $k_0 = \frac{2\pi}{\lambda}$ is the free-space wave vector, which corresponds in our case to the second harmonic wavelength. The phase distribution is then translated into the orientation angle of the C3 meta-atom by $\theta = \varphi/3\sigma$ (Supplementary Information S1). We designed the metalens with a diameter of 300 µm, whereas the C3 meta-atoms are aligned in polar coordinates on concentric rings with a spacing of 500 nm.

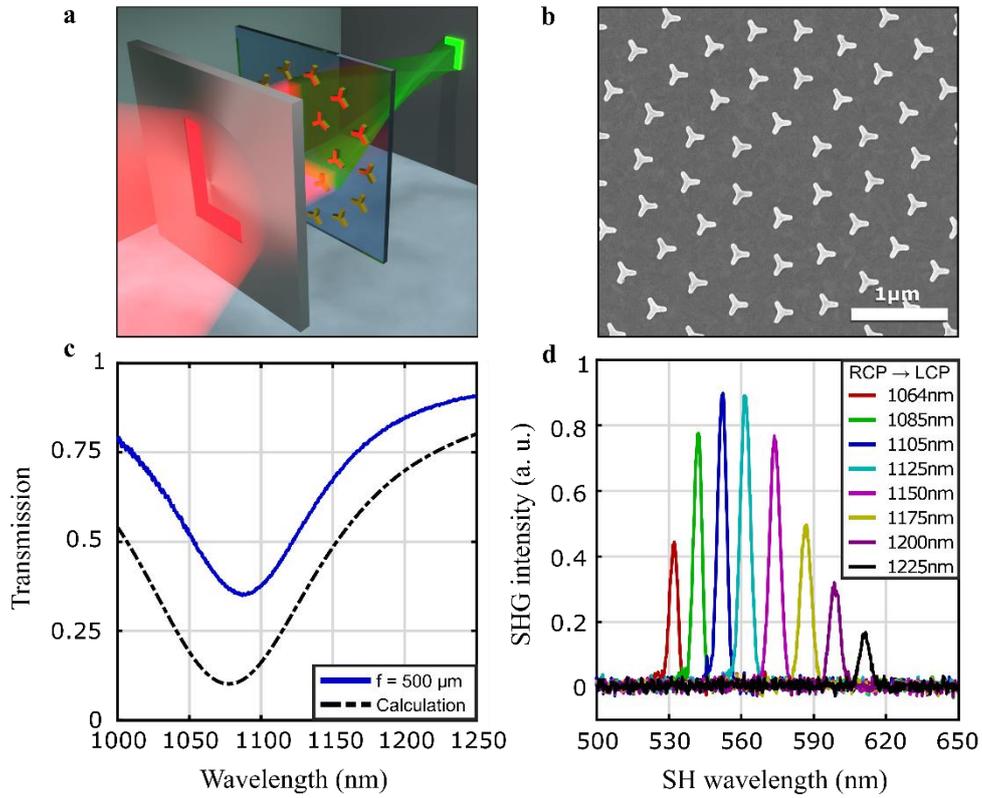

**Figure 1: Schematic concept and optical properties of the nonlinear metalenses. a** Illustration of the imaging concept: The L-shaped aperture is imaged on a screen with the help of the nonlinear metalens consisting of C3 nanoantennas. **b** Scanning electron microscopy image of the plasmonic metalens designed for a focal length of 500 µm. **c** Measured transmission spectra of the metalens. The simulation predicts a localized surface plasmon polariton resonance at a wavelength of 1064 nm, whereby the measured resonance dip is slightly shifted to 1085 nm. **d** Measured SHG signals for different right circularly polarized fundamental wavelengths with the maximum SHG intensity for RCP-LCP conversion between 1105 and 1125 nm. The SHG response disappears for RCP-RCP.

For our demonstration, we fabricate a set of different plasmonic metalenses with different focal length by using standard electron beam lithography (Figure 1b). For the experiments that are presented in this article we select the metalens with a focal length of $f = 500$ µm (for further results see the Supplementary Information S2).

First, we measure the linear transmission spectra of the lenses to determine the resonance wavelength (at approximately 1085 nm) of the localized plasmonic mode of the meta-atoms

(Figure 1c). Compared to the design wavelength of 1064 nm the resonance dips are slightly red-shifted, which is due to a small discrepancy of the fabricated meta-atom geometry from the target design (Supplementary Information S2). Although the dispersionless characteristics of the nonlinear P-B phase enable the SHG beam focusing and imaging across a broad range of wavelengths, the efficiency and the focal lengths vary for different wavelengths. The spectral response of SHG from the C3 metalens in forward direction is obtained for illumination with circularly polarized femtosecond laser pulses (Figure 1d). As expected, the SHG signal with the same polarization as that of the fundamental wave disappears as this process is forbidden by the nonlinear selection rule due to the C3 rotational symmetry of the meta-atoms.[31] In comparison, the SHG signal with cross-polarization shows a strong signal. We noted that the maximum SHG signal for a fundamental wavelength between 1105 nm and 1125 nm is stronger than that at the resonant wavelength of 1085 nm. This redshift is a result of the resonance that measures the wavelength dependent strength of the far-field enhancement, while the nonlinear conversion is sensitive to the near-field enhancement.[41] From the measured SHG signal strength we estimate a conversion efficiency in the order of $10^{-12}$ for our metalens design. Much larger values might be attainable by combining the plasmonic metalens with highly nonlinear materials.[34, 42]

Next, we experimentally investigate the focusing behavior of the metalenses for the second harmonic signal in forward direction. Note that there is a comparable SHG emission in backward direction, which was not investigated in our measurements. In our experiment, a circularly polarized fundamental Gaussian beam at different wavelengths is slightly focused onto the metasurface by a regular lens with focal length of $f = 500$ mm. The beam waist at the metalens position is marginally larger than the diameter of the metalenses to ensure that the full numerical aperture of the lens is used. The SHG signal from the metasurface is collected in the forward direction with an infinity corrected microscope objective (10 × / 0.25). The collected light is subsequently filtered using a short-pass filter. A tube lens with a focal length of $f_T = 200$ mm images the collected SHG signal onto a low noise CMOS camera (Andor Zyla). For analyzing the polarization state of the SHG we use a quarter waveplate (QWP) and a linear polarizer.

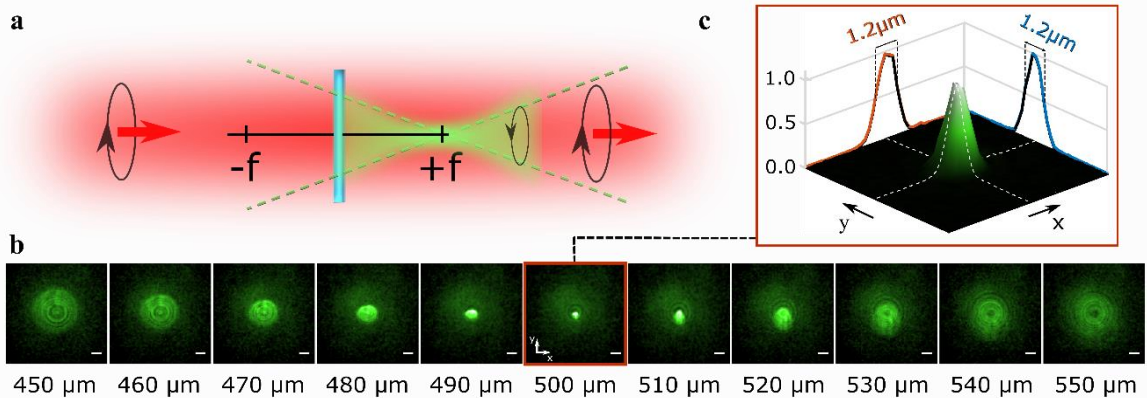

**Figure 2: SHG Measurement of the beam propagation in focal region. a** Schematic of the measurement: We use right circularly polarized incident beam with a wavelength of 1085 nm and a power of 75 mW, the converted left circularly polarized SHG signal is measured in the real focal region while the fundamental wave remains unaffected. **b** Measured SHG intensity distribution for the nonlinear metalens with f = 500 µm at different transverse planes in the focal region. The white scale bars correspond to a length of 10 µm and the SHG is encoded in false color on a logarithmic scale for better visibility. **c** Interpolated surface plot of the SHG focal spot on an area of 75 µm². The focal spot size for this metalens is 1.2 µm.

By changing the position of the microscope objective along the optical axis, different planes perpendicular to the propagation direction are imaged onto the camera. For the fundamental wave with RCP state, the metalens is designed to work as a convex lens with positive focal length for the SHG signal while the fundamental wave remains unaffected (Figure 2a). Further experiments with other focal lengths can be found in the Supplementary Information S4. For the fundamental wavelength of 1085 nm, the SHG is clearly focused at the designed focal length of 500 µm (Figure 2b). As for all our measurements, the point of reference for the beam propagation on the optical axis is the position of the nonlinear metalens ($z = 0$). We measure the Gaussian FWHM, which is 1.2 µm for each direction. To characterize the precise spot size, this measurement was done with a different microscope objective (40 × / 0.60) and a tube lens with a focal length of $f_T = 150$ mm. Reversing the circular polarization state of the fundamental wave leads to a concave phase profile with negative focal length for the SHG wave. In further measurements of the SH focusing, we verify the dual polarity of the nonlinear metalens by showing the real and virtual focal planes in dependence on different wavelengths (see Supplementary Information S4).

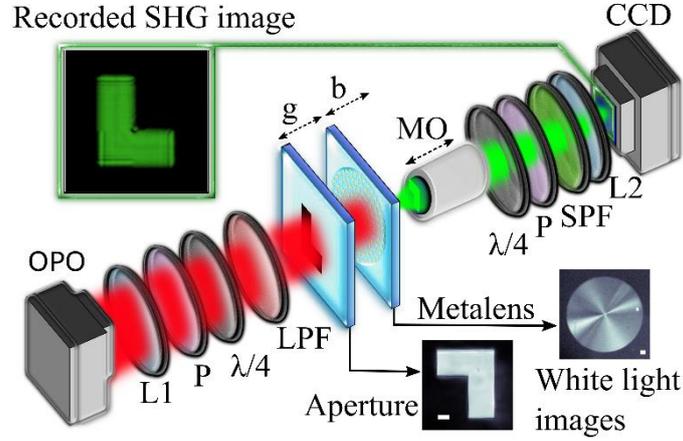

**Figure 3: Schematic illustration of the measurement setup.** For the nonlinear imaging, we illuminated different kinds of slot apertures with the fundamental beam. The insets show microscopy images of the L-shaped chromium aperture (thickness 100 nm) and of the metalens as well as the simulated SHG image (false color) at the camera position. λ/4, quarter waveplate; CCD, camera; L, lens; LPF, long-pass filter; OPO, optical parametric oscillator; P, polarizer; SPF, short-pass filter; MO, microscope objective lens. The scale bars correspond to 20 µm.

For demonstrating the imaging abilities of the nonlinear metalens, we fabricate an L-shaped aperture object as well as a double slit aperture (see Supplementary Information S5). We use these apertures as bright objects for the fundamental wave to provide sufficient power for the nonlinear process. The schematic of the imaging setup at the example for the L-shaped aperture is shown in Figure 3. When we illuminate the aperture by a right circularly polarized Gaussian beam at 1085 nm wavelength, we expect the metalens to work as a convex lens generating a real image of the aperture. First we place the aperture at the object distance $g = -f$ in front of the metalens. By adjusting the imaging plane of our microscope objective, we can retrieve the inverted SHG image of the L-shaped aperture with the magnification $\Gamma = -1$ at a certain image distance $b = 2f$ to the metalens. Note that the imaging system to the camera, which consists of the objective and the tube lens, also inverts the image, hence the real image after the nonlinear metalens is upright (in contrast to the white light image). Compared to the linear optical case, where the object and image distance is related by $-g = b = 2f$, we find that the nonlinear image formation is not governed by this relation. Hence, the traditional lens equation from linear optics cannot be applied to the nonlinear imaging case. To further study the nonlinear imaging properties of the metalens, we use a beam propagation method in the nonlinear regime (see Supplementary Information S3). With the help of this method, we are able to simulate the nonlinear evolution of the real and virtual image formation for our experimental conditions. We find that a one-to-one imaging can be realized with an object distance $g = -f$ and an image distance $b = 2f$. Interestingly, in linear optics this configuration would result in an image formation at infinity because the image distance will diverge. Based on the predictions of the numerical method we experimentally studied the nonlinear imaging progress in detail by recording the SHG images on the camera while we incrementally increase the distance to the metalens (Figure 4a).

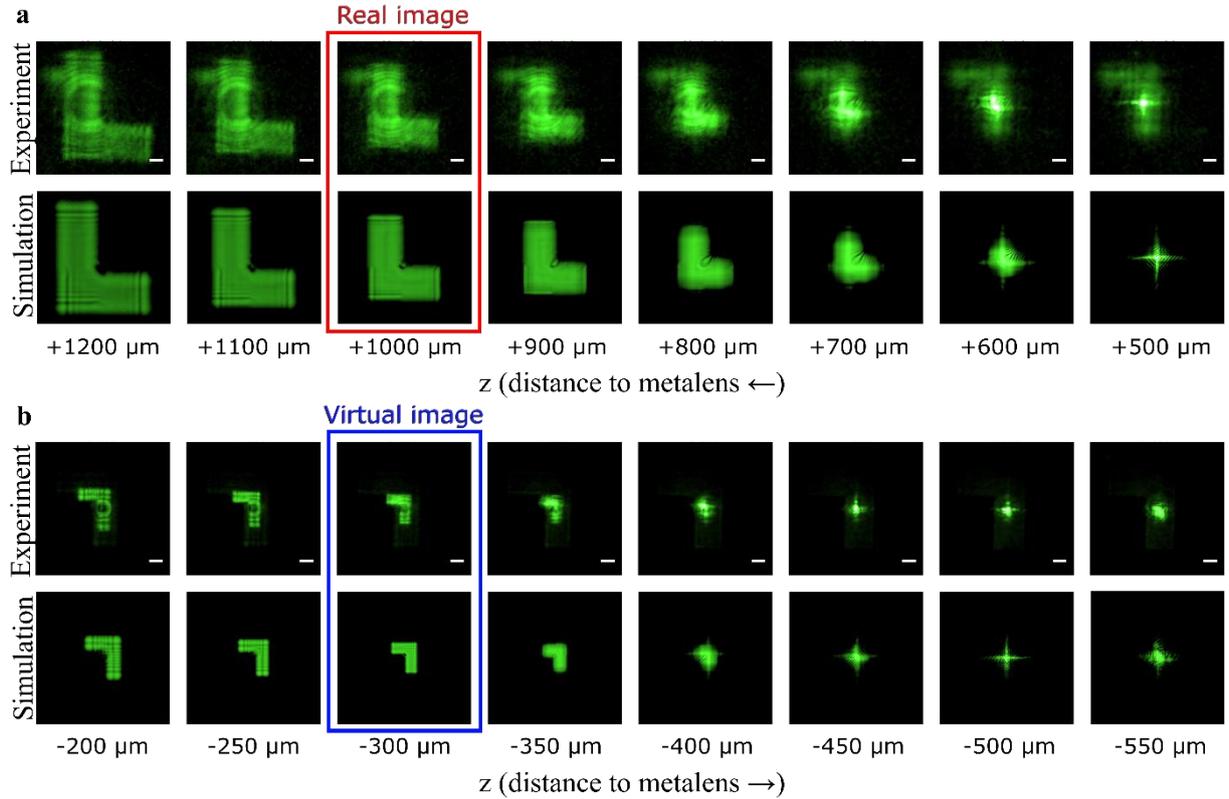

**Figure 4: 1:1 spin-dependent object imaging with nonlinear metalens. a** First row: Measured SHG distribution for planes along the z-propagation direction behind the metalens showing the evolution of the image formation. A clear upright real image of the letter L is formed at $z \approx 1000 \, \mu m$. Second row: Corresponding Simulation for the SHG distribution along the z-propagation based on beam propagation method in nonlinear regime. The metalens works as a convex lens for a spin conversion from RCP to LCP so that the real image is formed at $b = 1000 \, \mu m$ behind the metalens with the magnification $\Gamma = -1$. **b** For the reversed spin state of the fundamental beam (LCP) the nonlinear metalens acts as a concave lens which images an upside down virtual image of the letter L. The simulation shows a clear image 300 μm before the metalens. The measured image distance is $b = -300 \, \mu m$ with the measured magnification $\Gamma = 0.35$. White scale bars correspond to a length of $20 \, \mu m$. Each single transverse plane is $200 \times 200 \, \mu m^2$ in size and the SHG is encoded in false color on a logarithmic scale for better visibility. Note that all images are reversed by the additional imaging system.

For the RCP illumination ($f = 500$ µm), we measure the one-to-one mapping of the upright L-shaped aperture at an image distance of $b \approx 1000$ µm. The simulation (second row) is in good agreement with the measurement. Note that the interference fringes and the faint upside down L-shaped aperture image in the background come from scattering and refraction effects of the fundamental wave on the metalens and the aperture.

These effects disappear for the virtual image formation when the spin state of the fundamental wave is flipped from RCP to LCP and the sign of the focal length is altered, leading to an inverted virtual image of the L located between the L-shaped aperture and the metalens (Figure 4b). In our measurement the virtual image for fundamental LCP light appears at a distance of $b \approx -300$ µm with a magnification of $\Gamma = 0.35$. These values are close to the simulation results which show an image distance of $b \approx -333$ µm and a magnification of $\Gamma = 0.33$. Taking these results into account, we can empirically describe the image formation for small objects in small distances to the nonlinear metalens approximately by

$$\frac{1}{f} = \frac{1}{b} - \frac{1}{2g}, \tag{2}$$

whereas the magnification is given by

$$\Gamma = \frac{b}{2g}. \tag{3}$$

For further confirmation, we investigate the imaging of a larger T-shaped aperture and vary the object distance. All measurements are in good agreement with this approximation (see Supplementary Information S5). We also perform additional simulations for the image formation with THG and we find that the image equation can be generalized. Thus, the additional factor in front of our object distance corresponds to the harmonic generation order.

However, compared to a regular lens in the linear regime the nonlinear conversion process at the metalens may affect the image formation. To test the limitations of our nonlinear imaging metalens, we measured different double slit apertures with small gap distances to determine the influence of diffraction effects on the image quality. Due to the small feature size, the field distribution of the fundamental wave at the lens exhibits a typical far-field diffraction pattern. For the measurement, we use an object distance of $g = -4f$, which is more than 30 times larger than the slit distance between the two slits. A sharp real image with the highest contrast is measured in the SHG signal at a distance of 560 µm with $\Gamma = -0.12$ behind the metalens (Figure 5). The measured values confirm the results from our simulation, which predicts very similar image distance and magnification. However, the SHG image contains a weak third line between the images of the two slits. This additional feature in the image is the result of the nonlinear process at the metalens. We find that the squaring of the electromagnetic field during the SHG is affecting apertures more, which create an uncompleted interference pattern at the metalens position. Therefore, this imaging limitation depends on the object distance and on the harmonic generation order. In this context, one may interpret this as a new type of nonlinear image aberration. The effect can be obtained clearly for the measurement of the virtual SHG image with $\Gamma = 0.11$. However, such distortions can be reduced by decreasing the object distance or using double slit apertures with larger gap distances. Both would result in less diffraction of the object pattern.

Despite the distortion of imaging due to nonlinear interaction at the metasurface, the nonlinear metalens might have prospective advantages in the subject of nonlinear optical information processing.[43, 44] Indeed, the above experiment of the nonlinear imaging of the double slits already serves as a demonstration of a 'AND' logic gate if one considers the beams transmitting through the double slits as two input signals, and the weak central line of the nonlinear image as the output signal. The central line is present only if both the signals from the two slits interact nonlinearly at the metasurface, whereas blocking either one or both of the two slits would result in disappearance of the central line. Hence, such nonlinear metalenses might open new possibilities for performing nonlinear mathematical operations (like nonlinear Fourier transform) of optical signals.

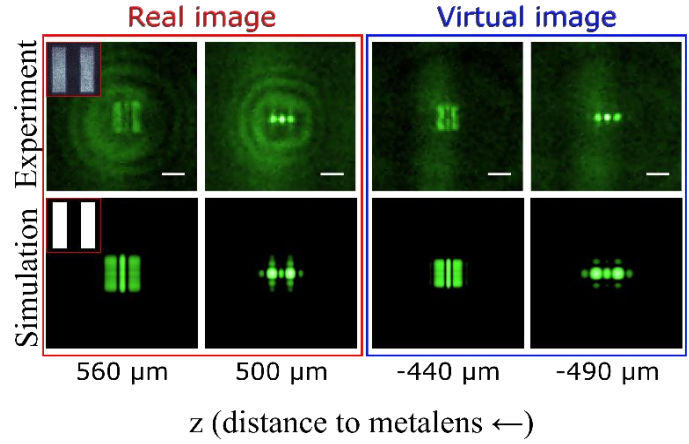

**Figure 5: Nonlinear imaging of double slit apertures.** The SHG real image of a double slit aperture with a gap distance of 60 µm and a slit width of 30 µm is obtained 560 µm behind the metalens (red framed). Diffraction orders are observed in the focal distance region at $z = 500\ \mu m$. Reversing the spin state of the fundamental wave (blue framed) results in a virtual image plane at z = -440 µm as well as the visibility of the diffraction orders at the virtual focal plane at -490 µm. The second row shows the simulation via beam propagation method. The object distance is $g = -4f$ and scale bars correspond to 20 µm.

In conclusion, we have experimentally demonstrated an ultrathin nonlinear metalens using second harmonic generation from gold meta-atoms with three-fold rotational symmetry. The desired phase profile for the metalens was obtained by a nonlinear P-B phase that is governed by the meta-atom orientation angle and the spin state of the fundamental wave. For a near infrared Gaussian laser beam, we experimentally realized the spin-dependent focusing effect of SHG waves at both real and virtual focal planes. Furthermore, we showed that objects, which are illuminated by near infrared light, can be imaged at visible wavelength based on the SHG process at the metalens. The symmetry of the meta-atoms only provides the desired P-B phase for the nonlinear process leaving the fundamental light unaltered in its propagation. This is in contrast to lenses made of regular nonlinear materials where both beam propagations are altered. The concept of nonlinear metalens not only inspires new imaging technologies but also provides a novel platform for generating and modulating the nonlinear optical waves.

## Methods:

Nanofabrication: A poly-methyl-methacrylate (PMMA) resist layer is spin-coated on ITO coated glass substrate and baked on a hotplate at 150 °C for 2 minutes to remove the solvent. The meta-atoms are patterned in the resist by using standard electron beam lithography, subsequent deposition of 2 nm chromium as an adhesion layer and 30 nm gold, followed by a lift-off procedure using acetone. The C3 plasmonic meta-atom has three assembled nanorods, whereas the length and width of each nanorod are set to 130 nm and 50 nm, respectively. The aperture objects are fabricated with the help of standard electron lithography and etching into a 100-nm-thick chromium layer. The L-shaped aperture is 100 µm in width and 120 µm in height, the symmetric T structure is 180 µm in both directions and the one-dimensional grating has a length of 200 µm and a period of 10µm.

Experiment: The incident circularly polarized femtosecond (fs) laser beam from an optical parametric oscillator is generated by using a linear polarizer and a quarter-wave plate. Then the fs laser of the desired wavelength is focused by an achromatic condenser lens ($f = 500$ mm) to a spot with a diameter of $\approx 300$ μm. The laser repetition rate is 80 MHz, the pulse duration $\sim 200$ fs and the used average power for the nonlinear measurements in Figure 2 is $\sim 75$ mW. The metalens is illuminated at normal incidence and the SHG signal is collected by an infinity corrected microscope objective ($\times 10 / 0.25$). To measure the spin state of transmitted SHG signal, a second quarter wave plate and a polarizer behind the microscope objective is added. A tube lens ($f = 200$ mm) is used to image the nonlinear signal onto the ultra-low noise sCMOS camera (Andor Zyla 5.5). For the measurement of the SH focal spot size the microscope objective ($\times 40 / 0.60$) and the tube lens ($f = 150$ mm) were adjusted. For the imaging of the letter shaped apertures, the average power of the laser is set to $\sim 100$ mW. For the double slit apertures this value is increased to $\sim 250$ mW. Furthermore, a short-pass filter is inserted in front of the camera to block the fundamental wave. In the experiment, the nonlinear metalens and the chromium apertures are facing each other.

**Acknowledgements**
This work was financially supported by the Deutsche Forschungsgemeinschaft (Grant Nos. DFG TRR142/A05 and ZE953/7-1). G.L. would like to thank the support from China "Recruitment Program of Global Experts" and Peacock program of Shenzhen. S.Z. acknowledges the support from European Research Council consolidator grant (TOPOLOGICAL). This project has received funding from the European Research Council (ERC) under the European Union's Horizon 2020 research and innovation programme (grant agreement No 724306).